\documentclass[graybox]{svmult}

\usepackage{mathptmx}       
\usepackage{helvet}         
\usepackage{courier}        
\usepackage{type1cm}        
\usepackage{aas_macros}
\usepackage{makeidx}         
\usepackage{graphicx}        
\usepackage{multicol}        
\usepackage[bottom]{footmisc}

\makeindex             

\newcommand{\msun}{\mbox{M$_\odot$}}

\begin{document}

\title*{A galactic-scale origin for stellar clustering}
\author{J.~M.~Diederik~Kruijssen}
\institute{J.~M.~Diederik~Kruijssen \at Max-Planck Institut f\"{u}r Astrophysik, Karl-Schwarzschild-Stra\ss e 1, 85748 Garching, Germany, \email{kruijssen@mpa-garching.mpg.de}}
%
%
\maketitle

\abstract*{We recently presented a model for the cluster formation efficiency (CFE), i.e. the fraction of star formation occurring in bound stellar clusters. It utilizes the idea that the formation of stars and stellar clusters occurs across a continuous spectrum of ISM densities. Bound stellar clusters naturally arise from the high-density end of this density spectrum. Due to short free-fall times, these high-density regions can achieve high star formation efficiencies (SFEs) and can be unaffected by gas expulsion. Lower-density regions remain gas-rich and substructured, and are unbound upon gas expulsion. The model enables the CFE to be calculated using galactic-scale observables. I present a brief summary of the model physics, assumptions and caveats, and show that it agrees well with observations. Fortran and IDL routines for calculating the CFE are publicly available at http://www.mpa-garching.mpg.de/cfe.}

\abstract{We recently presented a model for the cluster formation efficiency (CFE), i.e. the fraction of star formation occurring in bound stellar clusters. It utilizes the idea that the formation of stars and stellar clusters occurs across a continuous spectrum of ISM densities. Bound stellar clusters naturally arise from the high-density end of this density spectrum. Due to short free-fall times, these high-density regions can achieve high star formation efficiencies (SFEs) and can be unaffected by gas expulsion. Lower-density regions remain gas-rich and substructured, and are unbound upon gas expulsion. The model enables the CFE to be calculated using galactic-scale observables. I present a brief summary of the model physics, assumptions and caveats, and show that it agrees well with observations. Fortran and IDL routines for calculating the CFE are publicly available at http://www.mpa-garching.mpg.de/cfe.}

\section{The clustered nature of star formation} \label{sec:intro}
Studies of star formation in the solar neighbourhood, in the Milky Way as a whole, and in external galaxies, have shown that some fraction of star formation occurs in unbound associations, while the remainder results in bound stellar clusters (e.g. \cite{portegieszwart10,bressert12}). The question thus arises which physical mechanisms drive the star formation process to result in either outcome. How do bound stellar clusters form, and what fraction of all star formation do they represent? Observations suggest a range of answers to the latter question -- it seems that the fraction of star formation occurring in bound stellar clusters changes with the galactic environment \cite{goddard10,adamo11,silvavilla11,cook12}. This provides a potentially useful clue to the origins of stellar clustering, and presents a challenge to the classical view in which all stars form in clusters but only a minor fraction remains bound after gas expulsion (e.g. \cite{hills80}).

Another important insight is that star formation occurs over a broad and continuous range of densities \cite{bressert10}, without any obvious separation between star formation in unbound associations or bound clusters. Instead, protostellar cores and young stellar objects follow the hierarchical structure of the interstellar medium (ISM) \cite{efremov98}. It has recently been shown that gas-poor and virialised stellar structure may arise naturally at the high-density end of this density spectrum \cite{elmegreen08,kruijssen12}. This may provide a natural explanation to the observation that cluster formation is inefficient \cite{lada03}: if only some fraction of the star-forming regions manages to collapse and form bound systems, that same fraction will emerge as unembedded, virialized and bound clusters without any sign of expansion due to gas expulsion (e.g. \cite{cottaar12}).

In \cite{kruijssen12d}, we recently presented a new theoretical framework for the formation of bound stellar clusters. By integrating the star formation efficiencies (SFEs) and local bound fractions of star-forming regions over the density spectrum of the ISM, our model is used to quantify the fraction of all star formation that occurs in bound stellar clusters (i.e. the cluster formation efficiency or CFE). Fortran and IDL routines for calculating the CFE as a function of galaxy properties are publicly available at http://www.mpa-garching.mpg.de/cfe \cite{kruijssen12d}. In these proceedings, I summarize the model and discuss the underlying assumptions and caveats.

\section{The fraction of star formation occurring in bound clusters} \label{sec:model}
The theory of the CFE is derived and applied in detail in \cite{kruijssen12d}. In summary, it covers the following physical mechanisms and underlying assumptions.
\begin{enumerate}
\item{The starting point of the model is the overdensity probability distribution function (PDF) of the ISM. This PDF is assumed to follow a log-normal with median and dispersion set by the Mach number, and describes the distribution of density contrasts with respect to the mean density in a turbulent ISM (see e.g. \cite{krumholz05,padoan11}).}
\item{The overdensity PDF is written as a function of galaxy properties (gas surface density $\Sigma_{\rm g}$, angular velocity $\Omega$, and Toomre $Q$ \cite{toomre64}) by assuming that star formation occurs in a gas disc that obeys hydrostatic equilibrium (see below).}
\item{At each density the local SFE is calculated by assuming that the fraction of the gas that is converted into stars per free-fall time is approximately constant \cite{elmegreen02,krumholz05}. Depending on the density, star formation continues until (1) the gas is exhausted, (2) pressure equilibrium is reached between the turbulent ISM and (supernova and/or radiative) feedback, or (3) the moment of evaluating the CFE.}
\item{The local SFE is related to the local fraction of stars that remains bound upon instantaneous gas expulsion using a numerical simulation turbulent fragmentation \cite{bonnell08}. Assuming that protostellar outflows do not unbind bound stellar clusters, this provides the naturally bound fraction of star formation at each density.}
\item{Following the Spitzer formalism for tidal shocks \cite{spitzer87}, the model includes tidal perturbations by density peaks in the star-forming environment (the cruel cradle effect, see \cite{kruijssen11,kruijssen12}), which destroy stellar structure below a certain environmentally dependent, critical density.}
\item{The CFE is obtained by integrating the naturally bound fraction of star formation over the density range of the PDF where structure survives the cruel cradle effect (reflecting bound cluster formation), and dividing it by the integral of the SFE over the entire density range of the PDF (reflecting all star formation).}
\end{enumerate}
Given these model components, the following caveats should be kept in mind.
\begin{enumerate}
\item{Magnetic fields are only included to first order by using the magnetic-to-thermal pressure ratio, which is specified with an optional model parameter. This changes the dispersion of the log-normal overdensity PDF \cite{padoan11}. Note that the default form of this PDF is already consistent with weak magnetic fields.}
\item{The assumption that star formation occurs in a gas disc that obeys hydrostatic equilibrium may not be consistent with starburst galaxies. However, the energy dissipation that is required to cool gas and form stars also drives the formation of a disc, and hence the spatial distribution of star-forming regions in a starburst should be expected to follow a disc-like morphology \cite{hopkins09b}.}
\item{Star formation in intermediate-density regions is halted by supernova (or alternatively radiative) feedback, which may not be appropriate. The efficiency of different feedback mechanisms has been extensively discussed in the literature, and likely varies with spatial scale or density. However, the description of feedback in this model satisfies its purpose of truncating star formation on a timescale that is broadly consistent with observations \cite{portegieszwart10}.}
\end{enumerate}

\begin{figure}[t]
\sidecaption[t]
\includegraphics[width=7.2cm]{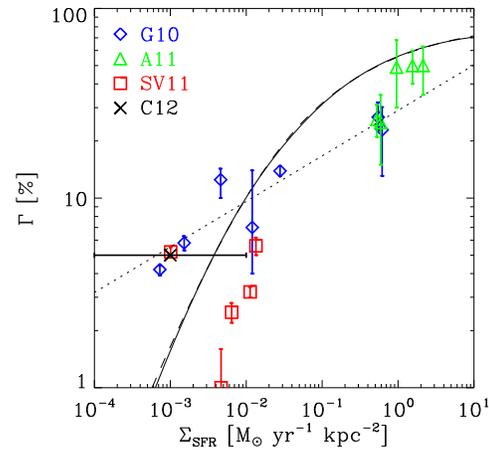}
\caption{CFE as a function of $\Sigma_{\rm SFR}$. Symbols denote observed galaxies with $1\sigma$ error bars and indicate the samples from \cite{goddard10} (blue diamonds), \cite{adamo11} (green triangles), and \cite{silvavilla11} (red squares). The black cross indicates the integrated CFE of all dwarf galaxies from the sample of \cite{cook12}, with a surface density range indicated by the horizontal error bar. The solid curve represents the modelled relation for typical disc galaxies, but can vary for different galaxy and ISM properties. The dashed curve shows the fit of equation~(\ref{eq:cfefit}), and the dotted line represents the original fit by \cite{goddard10}.}
\label{fig:cfesfrd}
\end{figure}

The above model can be used to calculate the CFE as a function of $\{\Sigma_{\rm g},\Omega,Q\}$, which is reduced to a single-parameter problem by assuming a single value of $Q=1.5$ and relating $\Omega$ to $\Sigma_{\rm g}$ as in \cite{krumholz05}. The resulting relation between the CFE (or $\Gamma$, \cite{bastian08}) and the star formation rate density $\Sigma_{\rm SFR}\propto\Sigma_{\rm g}^{1.4}$ \cite{kennicutt98b} is shown in Figure~\ref{fig:cfesfrd}, together with compiled observations from the recent literature. The agreement between theory and observations is remarkable, especially considering that the typical error margins on the observations are $\sim 0.3$~dex, and noting that an additional uncertainty of $0.3$--$0.5$~dex is introduced by using the relation $\Sigma_{\rm SFR}\propto\Sigma_{\rm g}^{1.4}$ to convert the model $\Gamma(\Sigma_{\rm g})$ to $\Gamma(\Sigma_{\rm SFR})$.
A good fit to the model relation is given by
\begin{equation}
\label{eq:cfefit}
\Gamma=\left(1.15+0.6\Sigma_{\rm SFR,0}^{-0.4}+0.05\Sigma_{\rm SFR,0}^{-1}\right)^{-1}\times100\% ,
\end{equation}
where $\Sigma_{\rm SFR,0}\equiv\Sigma_{\rm SFR}/\msun~{\rm yr}^{-1}~{\rm kpc}^{-2}$ is the star formation rate density of the galaxy. This fit is for one particular, `typical' parameter set and should therefore only be used for rough estimates. Because it assumes the power law form of the Schmidt-Kennicutt relation $\Sigma_{\rm SFR}\propto\Sigma_{\rm g}^{1.4}$ \cite{kennicutt98b}, any scatter around that relation is carried over. See \cite{kruijssen12d} and our publicly available routines for a more detailed modelling.

The good agreement between model and observations warrants further testing using Gaia and ALMA (see \cite{kruijssen12d} for an extensive discussion of the possibilities). However, our theoretical framework also contains several components that require to be constrained further. While a global theoretical picture of cluster formation seems to be emerging, the details of stellar clustering remain to be understood.

\begin{acknowledgement}
I am grateful to the organizers giving me the opportunity to present this work, and for organizing such a vibrant and pleasant conference.
\end{acknowledgement}

\bibliographystyle{spphys}

\end{document}